\documentclass[aps,prl,superscriptaddress,twocolumn]{revtex4}

\usepackage{bm}
\usepackage{color}
\usepackage{graphicx}
\usepackage{amsmath}
\usepackage{amssymb}

\begin{document}

\title{Tomographic measurement data of states that never existed}

\author{Julian G\"ottsch}
\affiliation{Institut f\"ur Quantenphysik \& Zentrum f\"ur Optische Quantentechnologien, Universit\"at Hamburg, Luruper Chaussee 149, 22761 Hamburg, Germany}
\author{Stephan Grebien}
\affiliation{Institut f\"ur Quantenphysik \& Zentrum f\"ur Optische Quantentechnologien, Universit\"at Hamburg, Luruper Chaussee 149, 22761 Hamburg, Germany}
\author{Felix Pein}
\affiliation{Institut f\"ur Quantenphysik \& Zentrum f\"ur Optische Quantentechnologien, Universit\"at Hamburg, Luruper Chaussee 149, 22761 Hamburg, Germany}
\author{Malte Lautzas}
\affiliation{Institut f\"ur Quantenphysik \& Zentrum f\"ur Optische Quantentechnologien, Universit\"at Hamburg, Luruper Chaussee 149, 22761 Hamburg, Germany}
\author{Daniela Abdelkhalek}
\affiliation{Institut f\"ur Quantenphysik \& Zentrum f\"ur Optische Quantentechnologien, Universit\"at Hamburg, Luruper Chaussee 149, 22761 Hamburg, Germany}
\author{Lorena Reb$\acute{\rm o}$n}
\affiliation{Institut f\"ur Quantenphysik \& Zentrum f\"ur Optische Quantentechnologien, Universit\"at Hamburg, Luruper Chaussee 149, 22761 Hamburg, Germany}
\affiliation{Instituto de F\'{i}sica La Plata, Universidad Nacional de La Plata, 1900 Buenos Aires, Argentina}
\author{Boris Hage}
\affiliation{Institut f\"ur Physik, Universit\"at Rostock, 18051 Rostock, Germany}
\author{Jarom\'{i}r Fiur\'{a}\v{s}ek}
\affiliation{Department of Optics, {Faculty of Science}, Palack\'y University, 17. listopadu 12, 77900 Olomouc, Czech Republic}
\author{Roman~Schnabel}
\email{roman.schnabel@uni-hamburg.de}
\affiliation{Institut f\"ur Quantenphysik \& Zentrum f\"ur Optische Quantentechnologien, Universit\"at Hamburg, Luruper Chaussee 149, 22761 Hamburg, Germany}

\date{\today}

\begin{abstract}
Microscopic Schr{\"o}dinger cat states are generated from quantum correlated fields using a probabilistic heralding photon subtraction event. Subsequent quantum state tomography provides complete information about the state with typical photon numbers of the order of one. Another approach strives for a larger number of quantum-correlated photons by conditioning the measurement analysis on events with exactly this number of photons.  
Here, we present a new approach to derive measurement data of quantum correlated states with average quantum-correlated photon numbers significantly larger than one. We produce an ensemble of a heralded, photon-subtracted squeezed vacuum state of light. We split the states at a balanced beam splitter and simultaneously measure a pair of orthogonal field quadratures at the outputs using tomographic `Q-function homodyne detection' (QHD). The final act is probabilistic two-copy data post-processing aiming for data from a new state with larger photon number. 
Evaluating the final tomographic data as that of a grown microscopic Schr{\"o}dinger cat state shows that the probabilistic post-processing increased the photon number of $|\alpha_0|^2 \approx 1.2$ to $|\alpha_2|^2 \approx 6.8$.  
Our concept for obtaining tomographic measurement data of mesoscopic non-classical states that never existed might be a turning point in measurement-based quantum technology. 
\end{abstract}

\maketitle

\vspace{-3mm}
\emph{Introduction} --- 
Stationary fields with quantum correlations and long coherence times are resources of quantum sensing \cite{LSC2011,Ganapathy2023}, quantum communication \cite{Gehring2015,Zhang2022}, and quantum computing \cite{Larsen2019,Fukui2022}.  
Generally speaking, the greater the (average) number of quantum-correlated photons $n_{\rm qc}$ in a state, the greater the quantum advantage.
For a squeezed vacuum state, the number of (quantum-correlated) photons can be derived from the measurable squeeze factor $\beta = \Delta^2 \hat X_{\rm vac} / \Delta^2 \hat X_\theta$, where $\Delta^2 \hat X_{\rm vac}$ is the variance of the ground state uncertainty of the electro-magnetic field and $\Delta^2 \hat X_\theta$ the variance of its most squeezed quadrature with $\theta$ being the squeeze angle. The number of quantum-correlated photons is then given by $n_{\rm qc} = (\beta + 1/\beta)/4 - 0.5$ \,\cite{Schnabel2017}. The highest measured squeeze factor so far is $\beta \!\approx\! 32$, which corresponds to approximately 15\,dB \cite{Vahlbruch2016} and $n_{\rm qc} \!\approx\! 7.4$ without any heralding signal or any other conditional detection event. Individual values of the field quadratures $X_{\rm vac,i}$ and $X_{\theta,i}$ are measured with balanced homodyne detection in a continuous fashion. Field quadratures of different angles (modulo $\pi$) do not commute. Orthogonal quadratures are often named $\hat X$ and $\hat Y$. They obey a Heisenberg uncertainty relation, and span a phase space for quasi-probability density distributions such as the Wigner function. The Wigner function is reconstructed from subsequent (tomographic) ensemble measurements of $\hat X$ and $\hat Y$, with one of them identical to $\hat X_\theta$ \cite{Vogel1989}. While squeezed states have an entirely positive Wigner function \cite{Breitenbach1997}, states with partially negative quasi-probability densities can be produced from them by the probabilistic subtraction of a single photon \cite{Dakna1997}. The subtraction event can be used as a heralding trigger that enables tomographic ensemble measurements of $\hat X$ and $\hat Y$. For weakly squeezed input states, the output state approximately corresponds to a microscopic Schr\"odinger cat states with average photon numbers of the order of one \cite{Ourjoumtsev2006,Neergaard-Nielsen2006,Ourjoumtsev2007,Asavanant2017}. 
For increasing $n_{\rm qc}$ of heralded microscopic Schr\"odinger cat-like states, the heralded generation followed by two-copy interference and (probabilistic) heralded growing was proposed \cite{Lund2004}. So far, only one growing step was realised \cite{Sychev2017,Konno2024}. In \cite{Sychev2017}, the size of a generated Schr\"odinger cat state increased from about 1.3 to about 3.4. Due to their detection efficiency of just 62\%, the actually measured size increased from $ |\alpha_0|^2 \!\approx\! 0.8$ to $|\alpha_1|^2 \!\approx\! 2.1$. Larger such states are the goal of highly active research in optical quantum computing \cite{Fukui2022}. 
Ref.
States with significantly larger numbers of quantum correlated photons were also produced, but heralding these states was not possible. The measurement of states with a deterministic photon number of a polarization entangled state of up to $n_{\rm qc} \!=\! 8$ were demonstrated \cite{Lu2007,Huang2011}. The states were conditional on the measurement itself, i.e.~on the $n_{\rm qc}$-fold coincidence of twice as many polarisation-filtered single-photon detectors. 

\begin{figure}[ht!!!!!!!!!!!!!!!!!!!!!!!!!!!!!!!!!!!!!!!!!!!!!!!!!!!!!!!!!]
\vspace{2mm} \hspace*{-0.5mm}
\includegraphics[width=0.92\linewidth]{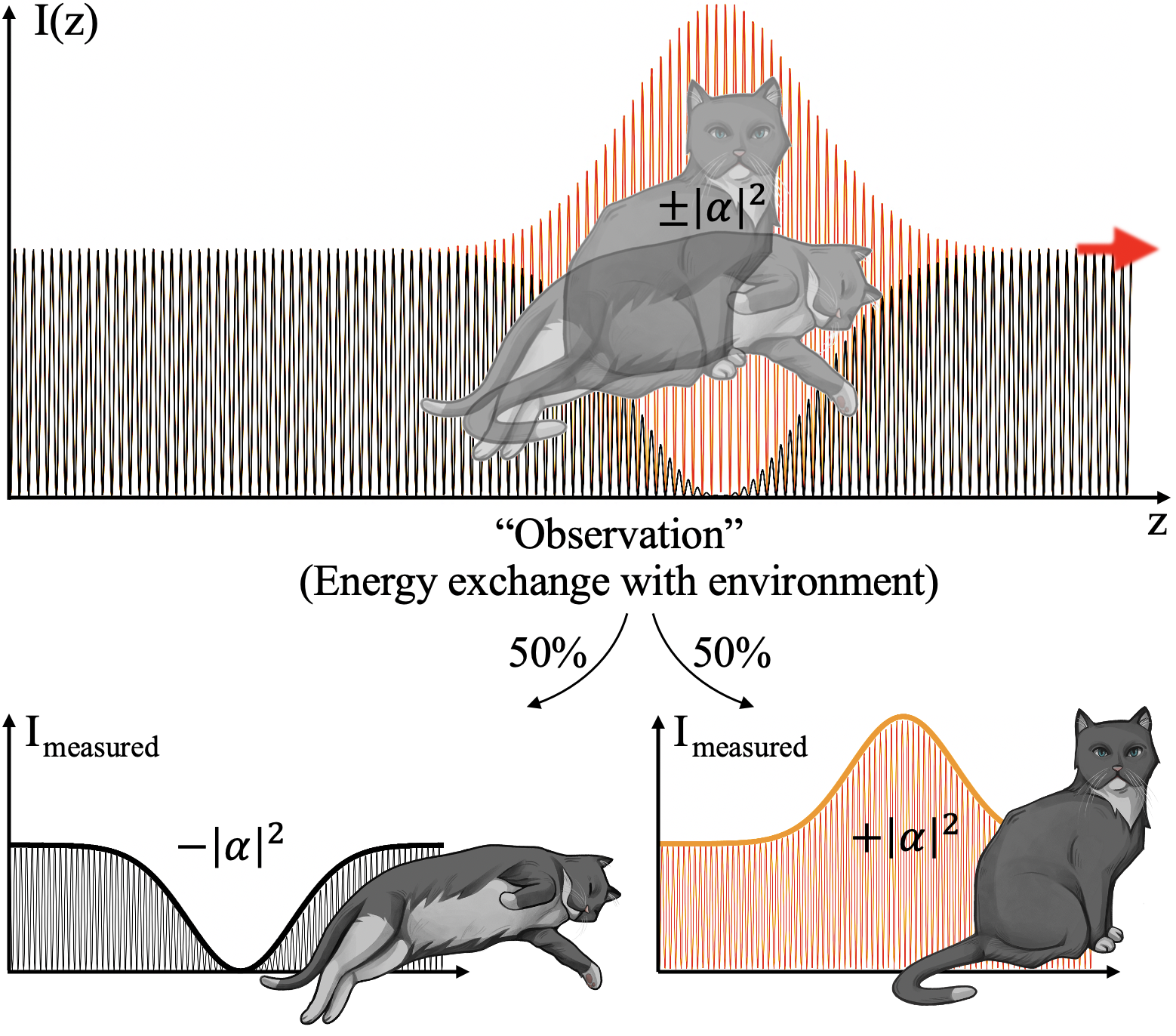}
\vspace{-2mm}
\caption{
{\bf Illustration of an optical Schr\"odinger cat} -- Optical wave packets (with a quantum correlated photon number $n_{\rm qc}$)
simultaneously constructively and destructively interfere with an optical local oscillator beam. The intensity measurement $I_{\rm measured}$ would show either a bright or a dark pulse with a `quantum-random' probability of 50\% each. Quantum randomness can be \emph{certified} by measuring  two non-commuting quadratures of the optical field  $\hat X$ and $\hat Y$. The tomographically reconstructed Wigner function proves a Schr\"odinger cat state including its macroscopicity $|\alpha|^2$ (brightness) and degradation due to decoherence. Different from previous works \cite{Ourjoumtsev2006,Neergaard-Nielsen2006,Ourjoumtsev2007,Asavanant2017}, we measured on each copy of the state the two non-commuting field strengths \emph{simultaneously} resulting in observables that we name $\hat X^Q$ and $\hat Y^Q$. 
}
\vspace{0mm}
\label{fig:1}
\end{figure}

Here, we report on a new experimental approach for the generation of tomographic measurement results on mesoscopic Schr{\"o}dinger-cat-like systems. We generated an ensemble of an heralded \emph{microscopic} Schr{\"o}dinger-cat-like states and performed tomographic measurements yielding $n_{\rm qc} = |\alpha_0|^2 \approx 1.2$ (without any correction). For this, a super-conducting nanowire detector subtracted a single photon from a weakly squeezed vacuum state followed by high-efficiency balanced homodyne detection, as achieved previously in Ref.~\cite{Asavanant2017}. 
Fig.\,\ref{fig:1} illustrates the concept of an optical Schr\"odinger cat state. 

The new part of our setup was the actual measuring device, a `Q-function homodyne detector', which was composed of two balanced homodyne detectors simultaneously measuring two orthogonal field quadratures ($\hat X^Q$ and $\hat Y^Q$) on every single state. `Q-function homodyne detection' (QHD) is known to provide the full ensemble information. We could thus apply data post-processing that combined the tomographic data of two states to probabilistically obtain new tomographic measurement data, as would had been obtained by measuring Schr{\"o}dinger-cat-like states with a `grown' photon number. 
After two growing steps, the data corresponded to that measured on a state with $|\alpha_2|^2 \approx 6.8$. Again, the number was not corrected for any inefficiencies in our setup. The generated `measurement data' were indistinguishable from actual measurement data on an ensemble of this state, although this ensemble did not physically exist as an optical ensemble at any time. Evidently, our approach required the complete measurement of the actually existing optical states, similar to the conditional measurement data generation in Ref.\,\cite{Lu2007,Huang2011}. Heralding the grown state was not possible.

\begin{figure}[t!!!!!!!!]
\centerline{\includegraphics[width=0.83\linewidth]{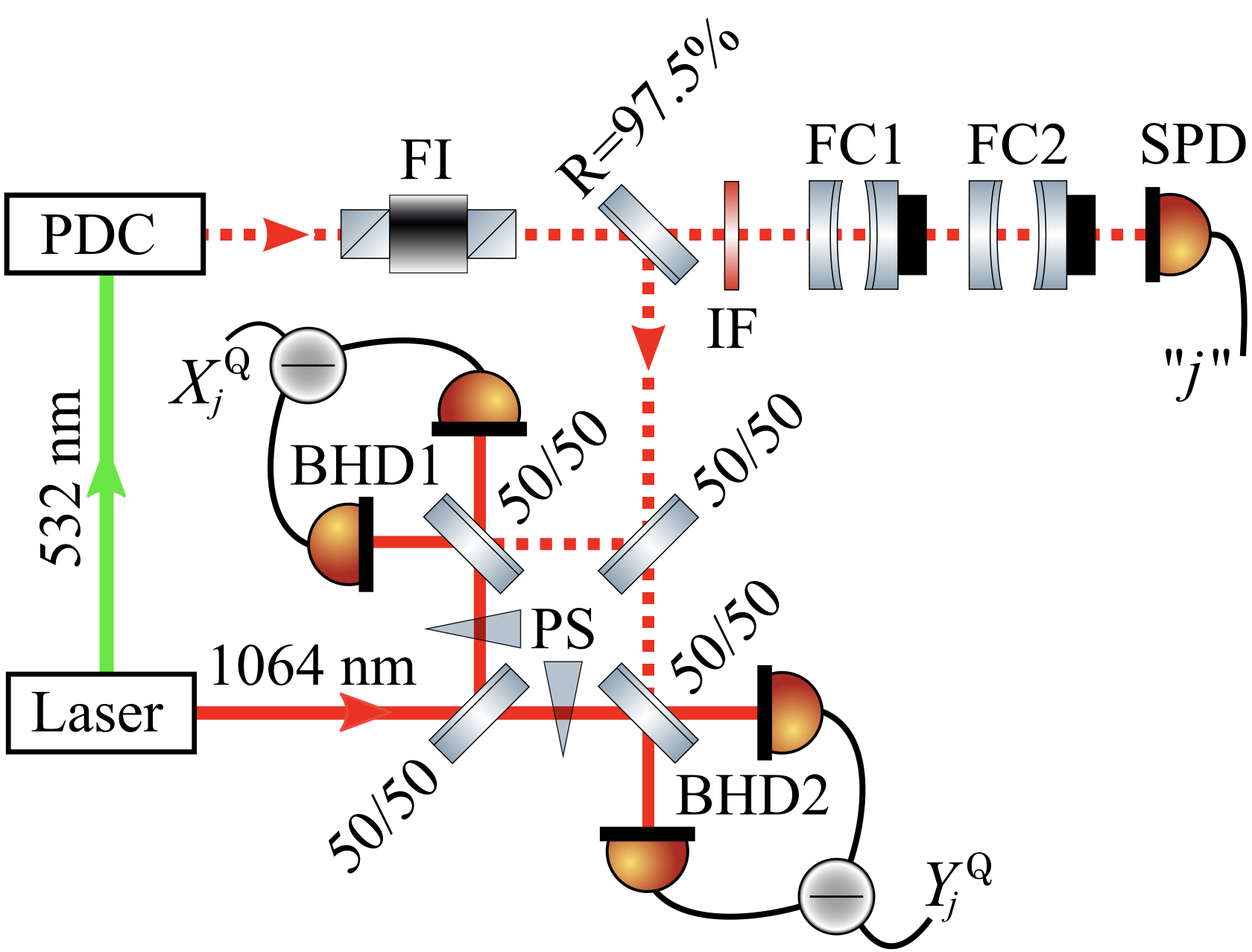}}
\vspace{-1mm}
\caption{
{\bf Laser interferometric setup} -- Light at 532\,nm pumps a parametric down-conversion process inside a crystal (PDC) to stationary produce states with squeezed uncertainty at 1064\,nm. Using interference with a 1064\,nm local oscillator, two balanced homodyne detectors (BHD1/2) continuously and simultaneously measure the field quadratures with the anti-squeezed electromagnetic field uncertainty $X^Q$ and with the squeezed one $Y^Q$ on 97.5\% of the states. If the single photon detector (SPD) registers a photon ``$j$'', the $X^Q_j$ and $Y^Q_j$ values in the corresponding time window belong to a microscopic Schr\"odinger cat state. The measured probability distribution of a typical ensemble is shown in Fig.\,\ref{fig:3} top left. FI: Faraday isolator; IF: interference filter; FC1, FC2: filter cavities of different lengths for matching the measurement spectrum of the SPD to that of the BHDs. PS: phase shifter.
}
\label{fig:2}
\end{figure}

\begin{figure*}[t!!!!!!!!!!!!!!!!!!!!!!!!]
\vspace{0mm}
\includegraphics[width=1\linewidth]{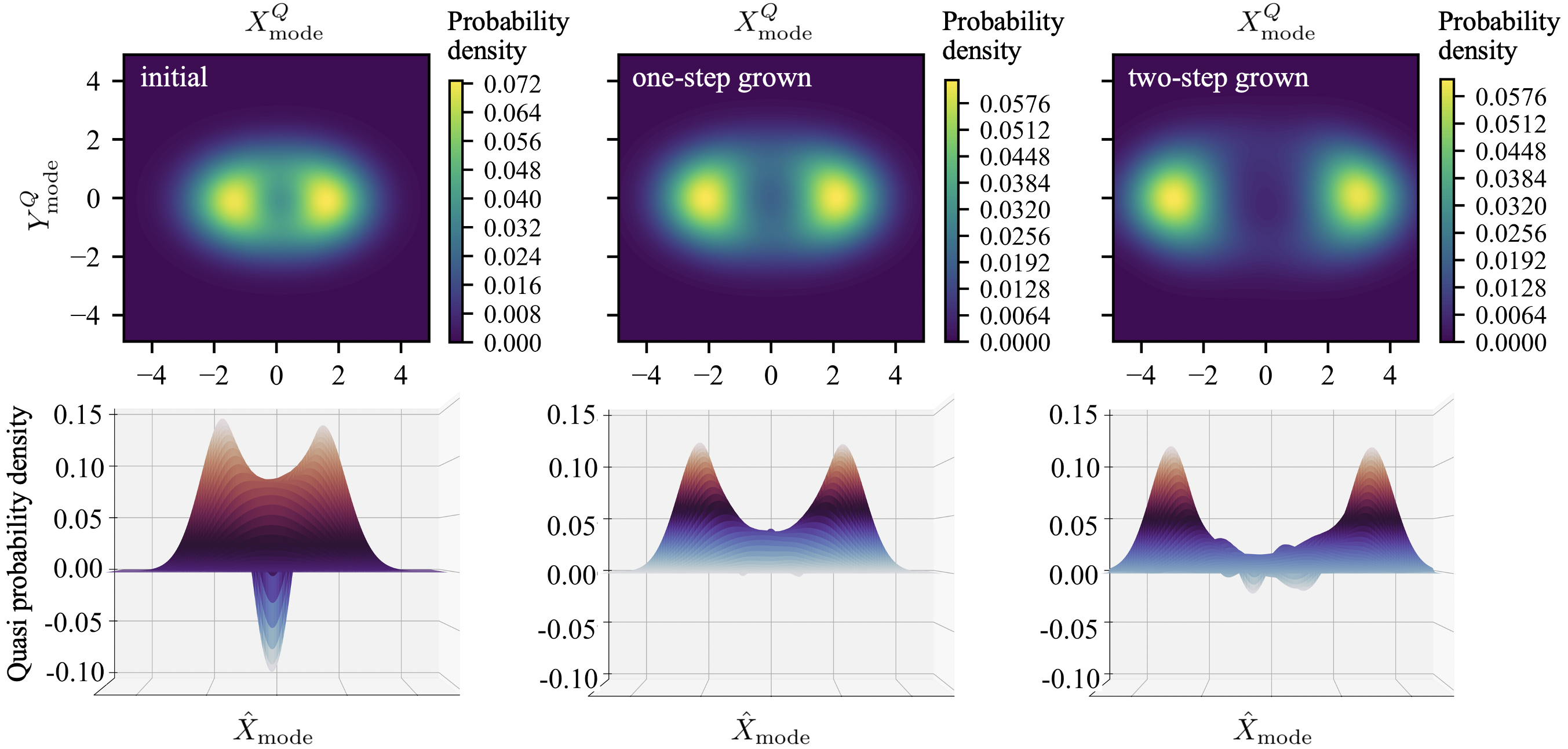}
\vspace{-3mm}
\caption{
{\bf Result of two growing steps without any corrections} -- Left column, top: Measured Q-function of the initial Schr\"odinger cat state $|\psi\rangle = (|\alpha_0 \rangle - |-\alpha_0 \rangle) / \sqrt{2}$ with $|\alpha_0| \approx 1.1$. Bottom: Quasi-probability distribution after deconvolution with the ground state uncertainty. The deeper the negativity, the more ideal the generation and detection of the state. Centre column: Q-function and Wigner function of the state after one growing step. Achieved was $|\alpha_1| \approx 1.7$. Right column: Q-function and Wigner function of the state after two growing steps. Achieved was $|\alpha_2|  \approx 2.6$. It's clearly to see that the post processing increased the amplitude $|\alpha|$ of the state since the maxima in Q and Wigner functions drifted apart.
}
\vspace{2mm}
\label{fig:3}
\end{figure*}
%

%
\emph{Experimental} --- Fig.\,\ref{fig:2} shows the schematic of our laser setup used for the generation of microscopic optical Schr\"odinger cat states, similarly to \cite{Neergaard-Nielsen2006}. Every such state was generated probabilistically, heralded by a detection event $j$ of a super-conducting nanowire single-photon detector (SPD) having a quantum efficiency of greater 93\% at 1064\,nm.
Altogether, we measured $1.2 \cdot 10^9$ time modes around heralding events being in the same microscopic Schr\"odinger cat-like state in about three and a half hours. We found a very good match with the theoretical superposition state of two opposing coherent states according to $|\psi\rangle_0 = (|\alpha_0 \rangle - |-\alpha_0 \rangle) / \sqrt{2}$ quantified by a fidelity of 62\% for $|\alpha_0| = 1.1$ without any corrections.

The innovation of our experiment was the \emph{simultaneous} measurement of two orthogonal, non-commuting field quadratures $X^Q_{\rm mode,j}$ and $Y^Q_{\rm mode,j}$, whose simultaneous precision is bounded by a Heisenberg uncertainty relation.
This type of detection has been named `8-port homodyne detection' or `heterodyne detection'. We introduce here the term `Q-function homodyne detection' (QHD) because we consider it more descriptive and less ambiguous than previous terms. It allows for the realistic simulation of measurement data from larger than actually available Schr\"odinger cat states. Before we come to this, we describe our result on the measurement of the initial microscopic Schr\"odinger cat states, where we do not apply any correction for e.g.~the imperfect quantum efficiency of the detection.

The ensemble of the pairs contained the full information about the Schr\"odinger cat state, which is conflated by the Husimi Q-function shown in Fig.\,\ref{fig:3} top left. The inset bottom left shows the corresponding Wigner function, which we calculated from the Q-function. 
This can be achieved by deconvolving the Q-function with the Gaussian function describing the ground state blur, but this procedure amplifies noise and other imperfections. To avoid direct deconvolution, we first reconstructed the density matrix in the Fock basis using the statistically well-motivated maximum likelihood reconstruction algorithm and then calculated the Wigner functions from it.
The Wigner function is the quasi-probability density function that describes the conventional ensemble measurement, where the ensemble is split into two to measure the two orthogonal quadratures separately, i.e.~without splitting the states.
The two representations on the left in Fig.\,\ref{fig:3} are clearly that of a Schr\"odinger cat state because they show two well-separated maxima with opposite signs along one of the field quadratures, and the Wigner function shows negativities in between.
The opposite sign describes the fact that the state simultaneously interferes constructively and destructively with a coherent state displaced along the same field quadrature, resembling Fig.\,\ref{fig:1}. The negativities prove that the measured state is not simply a mixed state of either a positive or a negative sign, and we can conclude that the representations in Fig.\,\ref{fig:3} left describe every single ensemble member. 
Note that the existence of local hidden variables that nevertheless would allow to divide the states into different subclasses were ruled out by the experimental violations of Bell inequalities \cite{Aspect1981,Giustina2013,Hensen2015}.

\emph{Generating measurement data on grown states} --- 
Measurement data from Q-function-homodyning can be post-processed to generate data from amplified states \cite{Fiurasek2012}. The advantage compared to a hardware-based approach \cite{Lund2004,Laghaout2013,Sychev2017} is groundbreaking. The data is indistinguishable, but neither doubling the quantum state generation hardware is required nor the integration of quantum memories. And most importantly, the success probability of this generally probabilistic approach is maximal, i.e.~as good as what is achievable by the integration of perfect quantum memories \cite{Abdelkhalek2016}.

We used the measured ensemble data of our microscopic Schr\"odinger cat state $|\alpha_0 \rangle$ and probabilistically created reduced ensemble data as measured on states $|\alpha_1 \rangle$ with $|\alpha_1| > |\alpha_0|$. 
The hardware-based growing uses overlapping of two copies of the state on a balanced beam splitter (BS) followed by the probabilistic generation of a trigger signal from one BS output port heralding the grown state in the other BS output port.
Here, we emulated this procedure by `overlapping' the measurement data from two ensemble members $|\alpha_0 \rangle_{j,j+1}$ on a virtual BS as follows
\begin{equation}
\alpha_{\rm j+}=\frac{1}{\sqrt{2}}(\alpha_{\rm 2j}+\alpha_{\rm 2j+1}), \qquad \alpha_{\rm j-}=\frac{1}{\sqrt{2}}(\alpha_{\rm 2j}-\alpha_{\rm 2j+1}) \, ,
\label{eq:1}
\end{equation}
were $\alpha_{\rm 2j} = X^{Q}_{\rm mode,2j} + i Y^Q_{\rm mode,2j}$ and $\alpha_{\rm 2j+1} = X^Q_{\rm mode,2j+1} + i Y^Q_{\rm mode,2j+1}$ 
with $1\!\leq\! j \!\leq\! 5 \!\times\! 10^6$.
If the amplitude in the destructively interfering output by chance obeyed $|\alpha_{\rm j-}|^2 < \bar{n}$, where $\bar{n}$ was a freely choosable hard boundary, measurement data on an enlarged state $|\alpha_1| > |\alpha_0|$ emerged in terms of $\alpha_{\rm j+} = X^Q_{\rm mode,\,j+} + i Y^Q_{\rm mode,\,j+} \equiv \alpha_1$.  

The centre column of Fig.\,\ref{fig:3} shows the ensemble result after one such growing step with $\bar{n} = 1.3$. 
The amplitude increased to $|\alpha_1| \approx 1.7$. Oscillations and negativities are still visible in the reconstructed Wigner function (centre, bottom), also the data is not corrected and imperfections increased due to the two-copy interference process.
The ensemble size is generally halved by the interference/growing process. On top, the growing process is probabilistic, and depending on the value $\bar{n} = 1.3$ only some fraction of the interferences is accepted as successful, in our case about 2\%, which finally resulted in a hundred-times reduced ensemble size.  
The right column of Fig.\,\ref{fig:3} shows the result of two-step grown measurement data according to Eq.\,\ref{eq:1}. The data corresponds to significantly decohered measurement data as it would be gathered from a Schr\"odinger cat state with  $|\alpha_2|  \!\approx\! 2.6$ ($\alpha^*_2 \alpha_2 \!\approx\! 6.8$) with a realistic, imperfect measurement setup.
It should also be noted here that the data is not corrected for any imperfections.
Further growth steps would be possible if a longer measurement period had provided a larger initial data set. Of course, the state decoherence would also increase further with each growth step.

%
\emph{Summary and conclusion} --- 
Our work demonstrates a new concept for generating conditional tomographic measurement data of states that cannot be generated physically. Here, the principle is proven on an initial ensemble of heralded, physically existing Schr\"odinger-cat-like states with the typical low photon number of the order of one. Simultaneous orthogonal field quadratures (Q-function value pairs) are measured on these states. The Wigner function reconstructed from the Q-function value pairs shows a highly significant negative quasi-probability density and corresponds to a typical microscopic Schr\"odinger cat state. Since the data correspond to the Q-function value pairs of individual states, the value pairs of two copies can be ``superimposed'' on a virtual beam splitter. The result at one output of the beam splitter must exceed a threshold value for the data at the output of the second beam splitter to improve, emulating a physical growth process as used in \cite{Sychev2017}. The outcome are emulated new Q-function value pairs measured on a grown state that never existed. The new ensemble data is indistinguishable from actual measured ensemble data. In our case, however, further growth steps can easily be added. The number of our iterative growth steps is only limited by the number of measured copies of the initial state, i.e.~by the measurement time. With each emulated growth step, the ensemble size is not only halved, but further reduced depending on the threshold(s). However, the probability of success is maximized, i.e.~as good as can be achieved by integrating \emph{perfect} quantum memories.\\
Since the grown state had never existed, it is impossible to herald it (before measuring the state). This precludes a general application of our concept in advanced quantum technologies, as is the case with other approaches to conditional generation of quantum states. However, our concept includes a variable data post-processing structure that can be easily adapted and optimized. We hypothesize that our concept of emulating measurement data of mesoscopic non-Gaussian states may pave the way for a subclass of quantum technology applications. 

\vspace{8mm}
\begin{acknowledgments}
\textbf{Acknowledgments} ---
This project is financed by ERDF of the European Union and by `Fonds of the Hamburg Ministry of Science, Research, Equalities and Districts (BWFGB)'. Until 2021, it was financed by the Deutsche Forschungsgemeinschaft (DFG) -- SCHN 757/7-1. L.R. was supported by the Consejo Nacional de Investigaciones Cient\'{i}ficas y T\'{e}cnicas (CONICET). J.F.  acknowledges support by the Czech Science Foundation under Grant No. 21-23120S. The dead and alive cats were produced by Kim-Melina Bertram.\\
\end{acknowledgments}

\textbf{Author Contributions} ---
D.A., J.F. and R.S. planned the experiment. J.G., F.P, M.L., S.G., D.A., L.R. and B.H. built and performed the experiment. F.P., S.G. and J.F. provided the theoretical analysis. S.G., J.F., and R.S. prepared the manuscript.\\

\textbf{Competing interests} ---
The authors declare no competing interests.\\

\textbf{Data availability} ---
The data that support the plots within this paper and other findings of this study are deposited into a public repository with the following accession code [will be available before publication].\\

\textbf{Code availability} ---
Our code, which we used to process our data for Fig.\,\ref{fig:3}, is stored in a public repository with the following access code [will be available before publication].\\ [-8mm]

\end{document}